\documentstyle[11pt]{article}

\def\kon#1#2{\vbox{\halign{##&&##\cr\lower4pt
\hbox{$\scriptscriptstyle\vert$}\hrulefill &\hrulefill\lower4pt
\hbox{$\scriptscriptstyle\vert$}\cr $#1$&$#2$\cr}}}

\def\ro{\varrho}

\def\d{\partial}
\def\=d{\,{\buildrel\rm def\over =}\,}

\def\sqr#1#2{{\vcenter{\vbox{\hrule height.#2pt\hbox{\vrule width.
#2pt height#1pt \kern#1pt \vrule width.#2pt}\hrule height.#2pt}}}}

\def\te{\vartheta}
\def\B{\Bigl}
\def\li{{\rm li}}

\begin{document}

\title{Nonstandard general relativity II}
\author{G\"unter Scharf \footnote{e-mail: scharf@physik.uzh.ch}
\\ Institut f\"ur Theoretische Physik, 
\\ Universit\"at Z\"urich, 
\\ Winterthurerstr. 190 , CH-8057 Z\"urich, Switzerland}

\date{}

\maketitle\vskip 3cm

\begin{abstract}  
We study Einstein's equations with matter in hydrostatic equilibrium in the nonstandard gauge which was recently investigated in the vacuum case. We obtain spherically symmetric solutions for any given rotation curve. These solutions can be used to describe the bulge of a galaxy without assuming dark matter.

\end{abstract}

\newpage

\section{Introduction}

What is nonstandard general relativity ? In standard GR one studies solutions of Einstein's equations up to equivalence classes under diffeomorphisms [1]. In [2] we have investigated the equivalence class of the Schwarzschild solution. Surprisingly enough we have found that different solutions in this class are physically inequivalent, that means observable quantities as the circular velocities are different. Consequently, for physics the classification by means of diffeomorphisms is too coarse; a more detailed analysis is required and this is the matter of nonstandard GR.
 
In [2] we have introduced a {\it physical} gauge fixing by requiring that an important observable, namely the circular velocity $V(r)$ has prescribed values. This leads to solutions completely different from Newtonian theory. We have studied the simplest situation of a static, spherically symmetric metric
$$ds^2=g_{\mu\nu}dx^\mu dx^\nu=e^a dt^2-e^bdr^2-r^2e^c(d\te^2+\sin^2\te d\phi^2)\eqno(1.1)$$
in spherical coordinates $(t,r,\te,\phi)$ which are defined by astronomical measuring techniques and we shall keep these coordinates throughout. The functions $a, b, c$ depend on $r$ only, and we take $c(r)\ne 0$ which is the essential difference to standard general relativity [3]. We have studied the vacuum solutions in [2]. Using them we have constructed disk-like solutions with a somewhat unphysical energy-momentum tensor. Here we consider the simplest physical energy-momentum tensor
$$t^\mu_\nu={\rm diag}(\ro,-p,-p,-p),\eqno(1.2)$$
which describes ordinary matter in hydrostatic equilibrium. Since $ds^2$ in (1.1) is assumed diagonal it makes no sense to consider more complicated $t_{\mu\nu}$ here.

In the next section we set up the corresponding field equations, and, taking energy-momentum conservation into account we find the same kind of algebraic dependence of the equations as in the vacuum case [2]: one of the three functions $a, b, c$ remains undetermined. We live with this and take again the circular velocity $V(r)$ as being given, to determine the metric completely. There results two coupled first order ordinary differential equations for a metric function and the pressure $p(r)$. In standard general relativity we have only one equation for $p(r)$ instead ([1], Sect. 11/1). In Sect.3 we solve these equations for a spherical bulge of a galaxy where the circular velocity is proportional to $r$ for small $r$. We find  the pressure which is finite and continuous at $r=0$. However, $p(r)$ cannot be expanded in powers of $r$ at $r=0$ because it involves the integral logarithm. In the last section we consider a slightly more general energy-momentum tensor with dust.
This gas plus dust problem has no solution in the static case.

\section{Einstein's equations in a spherically symmetric nonstandard gauge}

The Einstein tensor for the metric (1.1) has been calculated in [2]. Inserting the energy-momentum tensor on the right-hand side we obtain the following three field equations 
$$G_{tt}=e^{a-b}\Bigl[-c''-{3\over 4}c'^2+{1\over 2}b'c'+{1\over r}(b'-3c')\Bigl]+{1\over r^2}(e^{a-c}-e^{a-b})=
\kappa t_{tt}\eqno(2.1)$$
$$G_{rr}={1\over 2}a'c'+{1\over r}(a'+c')+{c'^2\over 4}+{1\over r^2}\Bigl(1-e^{b-c}\Bigl)=\kappa t_{rr}\eqno(2.2)$$
$$G_{\te\te}={r^2\over 2}e^{c-b}\Bigl[a''+c''-{1\over r}(b'-a'-2c')+{1\over 2}(a'^2-a'b'+a'c'-b'c'+c'^2)\Bigl]=
\kappa t_{\te\te},\eqno(2.3)$$
where $\kappa=8\pi G/c^4$ and $G$ is Newton's constant. The prime always means $\d/\d r$.  By combining (2.1) with (2.3) we write these equations in the form
$$c''=-{3\over 4}c'^2+{1\over 2}b'c'+{1\over r}(b'-3c')\Bigl]+{1\over r^2}(e^{b-c}-1)-\kappa t_{tt}e^{b-a}\eqno(2.4)$$
$$a''={c'^2\over 4}-{a\over 2}(a'-b'+c')+{1\over r}(c'-a')+{1\over r^2}(1-e^{b-c})+{2\kappa\over r^2}t_{\te\te} e^{b-c}+\kappa t_{tt}e^{b-a}.\eqno(2.5)$$
As in [2] we eliminate $b(r)$ using (2.2)
$$b=c+2\log r+\log\Bigl({1\over r^2}+{a'c'\over 2}+{c'^2\over 4}+{a'+c'\over r}-\kappa t_{rr}\Bigl)\eqno(2.6)$$
and
$$e^{b-c}=1+r^2\Bigl({a'c'\over 2}+{c'^2\over 4}\Bigl)+r(a'+c')-\kappa r^2t_{rr}\equiv M\eqno(2.7)$$
$$b'=c'+{2\over r}+{1\over 4M}\Bigl[a''(2r^2c'+4r)+c''2r^2(a'+c'+{2\over r})-4(a'+c')-{8\over r}-4\kappa r^2t'_{rr}\Bigl].
\eqno(2.8)$$
Then from (2.5) we get
$$c''={a''\over a'+c'+2/r}\Bigl(c'+{2\over r}+{4c'\over a'r}+{c'^2\over a'}+{4\over a'r^2}-4{\kappa\over a'}t_{rr}\Bigl)+$$
$$+{2M\over r^2}+{2\over r^2}+2\kappa{t'_{rr}\over a'+c'+2/r}-{4\kappa\over a'r^2}{M\over a'+c'+2/r}\Bigl[t_{rr}+
M\Bigl({2\over r^2}t_{\te\te}+t_{tt}e^{c-a}\Bigl).\eqno(2.9)$$
Substituting this into (2.8) we find
$$b'=2{a''\over a'}+a'+2c'+{4\over r}-2{\kappa\over a'}\Bigl[t_{rr}+M\Bigl({2\over r^2}t_{\te\te}+t_{tt}e^{c-a}\Bigl)\Bigl].
\eqno(2.10)$$
This allows to simplify (2.9) as follows
$$c''={a''\over a'}(c'+{2\over r})+{c'^2\over 2}+2{a'+c'\over r}+a'c'+{4\over r^2}+\kappa t_5\eqno(2.11)$$
where
$$t_5=-2t_{rr}-{a''\over a'+c'+2/r}{4\over a'}t_{rr}+{2t'_{rr}\over a'+c'+2/r}-{4M\over a'r^2(a'+c'+2/r)}t_4\eqno(2.12)$$
and
$$t_4=t_{rr}+M\Bigl({2\over r^2}t_{\te\te}+t_{tt}e^{c-a}\Bigl).\eqno(2.13)$$

Next we eliminate $b$ in (2.4) and find a result similar to (2.11):
$$c''={a''\over a'}(c'+{2\over r})+{c'^2\over 2}+2{a'+c'\over r}+a'c'+{4\over r^2}-\kappa t_c\eqno(2.14)$$
with
$$t_c=t_{tt}e^{c-a}M\Bigl[1+{1\over a'}\Bigl(c'+{2\over r}\Bigl)\Bigl]+t_{rr}\Bigl[1+{1\over a'}\Bigl(c'+{2\over r}\Bigl)\Bigl]+$$
$$+t_{\te\te}{2M\over a'r^2}\Bigl(c'+{2\over r}\Bigl).\eqno(2.15)$$
To see whether this is identical with (2.11-13) we insert the components of the energy-momentum tensor (1.2). Lowering one index in (1.2) we have
$$t_{tt}=e^a\ro,\quad t_{rr}=e^bp,\quad t_{\te\te}=r^2e^cp.\eqno(2.16)$$
Then $t_c$ (2.15) is equal to
$$t_c=e^b\Bigl\{\ro\Bigl[1+{1\over a'}\Bigl(c'+{2\over r}\Bigl)\Bigl]+p\Bigl[1+{3\over a'}\Bigl(c'+{2\over r}\Bigl)\Bigl]
\Bigl\}.\eqno(2.17)$$
On the other hand $t_5$ (2.12) contains the derivative of $t_{rr}$. This is the point where energy-momentum conservation comes in. In hydrostatic equilibrium we have ([1], eq.(5.4.5))
$$\d_r p=-(p+\ro)\d_r\log g^{1/2}_{tt},\eqno(2.18)$$
which gives
$$p'(r)=-(p+\ro){a'\over 2}.\eqno(2.19)$$
Now we are able to calculate
$$t_{rr}'=e^b(p'+pb')$$
$$=e^b\Bigl[-(p+\ro){a'\over 2}+p\Bigl( 2{a''\over a'}+a'+2c+{4\over r}-2{\kappa\over a'}t_4\Bigl)\Bigl]\eqno(2.20)$$
where (2.10) has been used. Substituting this into (2.12) we finally end up with the simple result
$$t_5=e^b\Bigl[-p-\ro-{f'\over a'}(\ro+3p)\Bigl].\eqno(2.21)$$
Here we have introduced the function
$$f'=c'+{2\over r}\eqno(2.22)$$
or
$$f(r)=c(r)+2\log{r\over r_c},\eqno(2.23)$$
which was already used in [2]; $r_c$ is a constant of integration. Comparing (2.21) with (2.17) we see that $t_5=-t_c$. Consequently, the two field equations (2.11) and (2.14) are identical which is the same degeneracy as in the vacuum case [2].

As in the vacuum case we lift the degeneracy by requiring that the circular velocity $V(r)$ has given values. The latter is equal to ([2], eq. (2.12))
$$V^2(r)\equiv u={a'\over f'}.\eqno(2.23)$$
This enables us to eliminate $a(r)$ in favor of $u(r)$ and $f(r)$.  From (2.19) we obtain the differential equation for the pressure
$$p'=-{u\over 2}(p+\ro)f'\eqno(2.24)$$
Next we rewrite (2.14) in terms of $f$ and $u$. It is a remarkable property that the second derivatives drop out, so that we get a first order differential equation for $f$:
$$f'^2(u+{1\over 2})+f'{u'\over u}=\kappa t_c.\eqno(2.25)$$
This equation is coupled to (2.24) because $t_c$ contains the pressure $p$. In standard general relativity on has only one first order differential equation for the pressure ([1], eq.(11.1.13)).

To work out $t_c$ in detail we start from
$$e^b=r^2e^c\Bigl({1\over r^2}+{a'c'\over 2}+{c'^2\over 4}+{a'+c'\over r}-\kappa t_{rr}\Bigl).$$
Inserting $t_{rr}=p\exp b$ and 
$$e^c=\B({r_c\over r}\B)^2e^f$$
we can solve for
$$e^b={r_c^2e^f\over 1+\kappa r^2e^fp}f'^2\B({u\over 2}+{1\over 4}\B).\eqno(2.26)$$
This finally leads to
$$t_c=\B(p+\ro+{f'\over a'}(3p+\ro)\B)e^b$$
$$=\B[p\B(1+{3\over u}\B)+\ro\B(1+{1\over u}\B)\B]\B({u\over 2}+{1\over 4}\B){r_c^2e^f\over 1+\kappa r^2e^fp}.\eqno(2.27)$$
Substituting this into (2.25) gives us a closed system of first order differential equations for $f(r)$ and $p(r)$. As in standard general relativity an equation of state $\ro=\ro(p)$ must also be known.

\section{Integration of the equations for small $r$}

Regarding the application of our spherically symmetric metric to real galaxies it is a pity that spherical galaxies are rare. If one analyzes disk galaxies, for example M33 [4] [5], one represents the stars in the disk by some density distribution which represents the measured surface brightness. However, then quite often there remains an emission excess in the central region. This is attributed to a nucleus or bulge, and it is not bad to describe this bulge by a nonstandard spherically symmetric metric and corresponding matter density and pressure. Therefore we want to solve our above equations for small $r$.

The behavior of the rotation velocity $V(r)$ for small $r$ is well known [6], $V(r)$ vanishes linearly for $r\to 0$,
therefore we assume
$$u(r)=u_2 r^2+O(r^3)\eqno(3.1)$$
where $u_2$ is a positive constant. Regarding the matter density $\ro$ we set $\ro ={\rm const}$ throughout to find the most simple solution. Then as the zeroth approximation we also put the pressure $p$ to be constant $p=p_0$, but later on we calculate corrections to this. Under these assumptions we find a singular behavior for $t_c$ (2.27)
$$t_c={t_2\over r^2}\eqno(3.2)$$
where $t_2$ is another positive constant. Now equation (2.25) assumes the simple form
$$r^2f'^2+4rf'=2\kappa t_2e^f.\eqno(3.3)$$
We solve the quadratic equation
$$r{df\over dr}=-2\pm\sqrt{4+2\kappa t_2e^f}\eqno(3.4)$$
and we integrate this equation by separating the variables
$${dr\over r}={df\over -2(1\mp\sqrt{1+{\kappa\over 2} t_2e^f}}.\eqno(3.5)$$

Only the upper minus sign in (3.5) leads to a physical solution. After integration we arrive at the equation
$$-2\log{r\over r_f}={1\over 2}\log\B\vert{1+y\over 1-y}\B\vert +{1\over y-1}\eqno(3.6)$$
where
$$y=\sqrt{1+{\kappa\over 2} t_2e^f}\approx 1+{\kappa\over 4}t_2e^f\eqno(3.7)$$
and $r_f$ is a constant of integration.
We shall soon see that $\exp f$ is small for small $r$ such that we can expand the square root in (3.7). For the same reason we can neglect the logarithmic term in (3.6) so that
$$-2\log{r\over r_f}={4\over\kappa t_2}e^{-f}\eqno(3.8)$$
or
$$e^f={2\over\kappa t_2\log\vert r/r_f\vert}\eqno(3.9)$$
for $r<r_f$. Indeed this goes to zero for $r\to 0$. Similarly from (3.4) we find
$$f'={2\over r}\B(-1+\sqrt{1+{\kappa\over 2}t_2e^f}\B)\approx {\kappa\over 2}t_2{e^f\over r}=$$
$$={1\over r\log\vert r/r_f\vert}.\eqno(3.10)$$
We also note
$$f(r)=\log{2\over\kappa t_2}-\log\log\B\vert{r\over r_f}\B\vert\eqno(3.11)$$
which follows from (3.9).

Next we must integrate the equation (2.24) for the pressure
$$p'=-{u_2\over 2}r^2(p+\ro)f'.\eqno(3.12)$$
Using
$$r^2f'={r\over\log\vert r/r_f}\vert$$
the solution of the homogeneous equation is given by
$$p(r)=p_1\exp\B[{u_2\over 4}r_f^2\li \B({r\over r_f}\B)^2\B]$$
where $p_1$ is a constant of integration and
$$\li z=\int\limits_0^z{dx\over\log x}\eqno(3.13)$$
is the integral logarithm. Then the solution $p(r)$ of (3.12) with $p(0)=p_0$ is equal to
$$p=(p_0+\ro)\exp\B[{u_2\over 4}r_f^2\li\B({r\over r_f})^2\B)\B]-\ro.\eqno(3.14)$$
The integral logarithm has no expansion in powers of $r$ at $r=0$, so we leave (3.14) as its stands. Still the pressure is continuous and differentiable at $r=0$.

We also want to calculate the metric, although it does not have a direct physical meaning. From (3.9) we immediately find
$$e^c=\B({r_c\over r}\B)^2e^f={2r_c^2\over\kappa t_2r^2\vert\log r/r_f\vert}.\eqno(3.15)$$
The $tt$-component follows from
$$a'=uf'={u_2r\over\vert\log r/r_f\vert}$$
which gives
$$e^a=K_a\exp\B({u_2\over 2}\li ({r\over r_f})^2\B).\eqno(3.16)$$
Finally, $\exp b$ is obtained from (2.26). In leading order we have
$$e^b={r_c^2\over 2\kappa t_2}{1\over r^2\vert\log r/r_f\vert^3}.\eqno(3.17)$$
So the metric has a singularity at $r=0$, but there is no horizon at some finite $r$ as in the Schwarzschild solution.

\section{The gas plus dust problem}

The energy-momentum tensor (1.2) which we have considered so far can be interpreted as coming from a gas in hydrostatic equilibrium. Now we want to add dust with density $\ro_0$ to this gas so that we assume a slightly more general
$$t^\mu_\nu={\rm diag}(\ro+\ro_0,-p,-p,-p).\eqno(4.1)$$
By definition the dust has no pressure; that means $p$ is the gas pressure as before. The additional $\ro_0$ violates energy-momentum conservation, therefore we expect some obstruction.

It is not hard to repeat the calculations in sect.2 with this more general energy-momentum tensor (4.1). Instead of (2.17) we have
$$t_c=e^b\B\{(\ro+\ro_0)\B(1+{f'\over a'}\B)+p\B(1+3{f'\over a'}\B)\B\},\eqno(4.2)$$
but the energy-momentum conservation (2.19) remains unchanged because it holds for the gas only. Regarding $t_5$, equation (2.12) still holds, but (2.13) becomes
$$t_4=e^b(\ro+\ro_0+3p).\eqno(4.3)$$
which causes a change in $t'_{rr}$ (2.20). Substituting all this into (2.12) we finally obtain
$$t_5=-t_c+\ro_0e^b{a'\over a'+f'}.\eqno(4.4)$$
Here the last term destroys the previous degeneracy in the Einstein's equations. Moreover one easily sees that there is no solution of the equations for $\ro_0\ne 0$.

This negative result has a clear physical interpretation: static dust is impossible. Dust is also used to represent stars in a galaxy model. Then the physics is pretty clear: the stars must move. This gives additional components in the energy-momentum tensor (4.1) and, as a consequence, the spherical symmetry is destroyed.

\end{document}